\documentclass[runningheads]{svmult}

\usepackage{makeidx}   
\usepackage{graphicx}  
\usepackage{subeqnar}  
\usepackage{multicol}  
\usepackage{physprbb}  
\makeindex             

\begin{document}
\title*{Optical spectra of dusty starbursts}
\toctitle{Optical spectra of dusty starbursts}
%
%
\titlerunning{Optical spectra of dusty starbursts}
%
\author{Bianca~M.~Poggianti\inst{1}}
\authorrunning{Bianca~M.~Poggianti}
%
%
\institute{Osservatorio Astronomico, Padova 35122, Italy}

\maketitle              

\vspace{-0.5in}
\begin{abstract}
This contribution presents the optical spectral properties 
of FIR-luminous galaxies, whose distinctive feature is often the
simultaneous presence in the spectra 
of a strong $\rm H\delta$ line in absorption
and of emission lines (e(a) spectra). 
A discrepancy between the star formation rate estimated from the
FIR luminosity and that derived from the $\rm H\alpha$ luminosity
persists even after having corrected the $\rm H\alpha$
flux for dust according to the observed Balmer decrement.
It is shown that the e(a) spectrum can be reproduced
assuming a current starburst and a dust extinction
affecting the youngest stellar populations much more
than the older stars.
\end{abstract}

\section{Introduction}
The most distinctive feature in the optical spectra of FIR-luminous
galaxies is not the equivalent width of the emission lines\footnote{These 
are usually lower than those in HII galaxies and UV-bright starbursts.}, 
but the equivalent 
width of the $\rm H\delta$ line in absorption, which is generally
stronger than in the spectra of optically selected galaxies.
In fact, emission line spectra with EW($\rm H\delta)>4$ \AA $\,$
(e(a) spectra) are frequent among FIR galaxies, while the great 
majority of nearby spirals in optical samples have EW($\rm H\delta)<4$ \AA $\,$
(e(c) spectra) (Poggianti et al. 1999 [P99]; Poggianti \& Wu 2000 [PW00]).
The e(a) spectral class differs from the so-called "k+a" (or "E+A")
galaxies which only display a strong $\rm H\delta$ line in absorption, but
no emission lines.

What is the origin of the difference in the $\rm H\delta$ strength
between dusty starburst galaxies and quiescent spirals?
What star formation and dust properties generate
the peculiar spectral combination found in e(a) galaxies?
In the following I will address these issues presenting the analysis
of a spectroscopic survey of luminous IRAS galaxies (PW00)
and the results of an effort to model the e(a) spectrum in detail
(Poggianti, Bressan \& Franceschini 2001).

\section{FIR-luminous galaxies: spectral features and star formation 
properties}

In PW00 we analyzed the spectral characteristics of
a complete sample of IRAS galaxies
(Wu et al. 1998a,b) comprising 73 Very Luminous Infrared galaxies 
(VLIRGs, $log(L_{IR}/L_{\odot})> 11.5$) with a median 
$log(L_{IR}/L_{\odot})= 11.72$
and 40 companion galaxies, selected from the 2 Jy
redshift survey of Strauss et al. (1992). The great majority  
of these galaxies show evidence for a strong interaction or a merger
(Wu et al. 1998b). The spectra typically cover the central 2 kpc of 
the galaxies at a $\sim 10$ \AA $\,$ resolution.

Table~1 shows the fraction of VLIRGs as a function of the spectral class;
a detailed description of each class can be found in PW00.
More than half of these galaxies have an e(a) spectrum, about 1 out of 4
has an e(c) spectrum, and only 1 out of 10 has very strong 
[O{\sc ii}] emission (e(b) type). None of the e(a) spectra would be
classified as a Seyfert1 or 2 according to the standard diagnostic diagrams
of line intensity ratios.
Notably no k+a spectrum is found among the FIR luminous galaxies, neither
among their companions: all the VLIRGs display emission lines in their spectra.

\begin{figure}[b]
\vspace{-1.9in}
\begin{center}
\includegraphics[width=.8\textwidth]{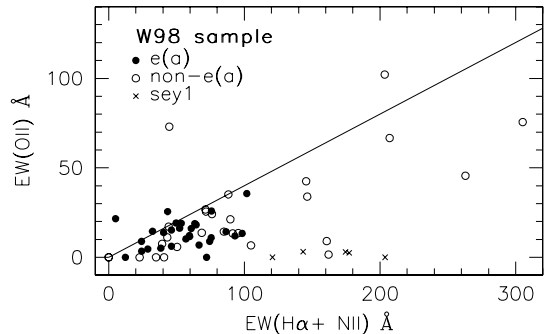}
\end{center}
\vspace{-0.6in}
\caption[]{Line represents the 
fit for normal field galaxies in the local Universe
(EW([O{\sc ii}])=0.4 EW($\rm H\alpha$ + NII)).}
\label{eps1}
\end{figure}
A general characteristic of the e(a) spectra in all environments and at all 
redshifts appears to be the low [O{\sc ii}]/$\rm H\alpha$ ratio (see Fig.~1): 
both the
equivalent width and the flux ratios of these lines are a factor of 2
lower than the median ratios observed in nearby spirals. Though 
such low ratios may be caused by various reasons, the most likely explanation
is reddening by dust: the color excess
E(B-V) derived from the observed Balmer decrement 
(see Table~1) is consistent with -- and fully accounts for -- 
the observed [O{\sc ii}]/$\rm H\alpha$ values.
It is noteworthy that the difference in the {\em equivalent width}
ratio between this sample and optically selected 
nearby spirals is entirely due to 
the difference in the average {\em flux} ratio of these two lines.
This can be interpreted as an effect of {\em selective dust extinction}
which affects the youngest stellar generations (still deeply embedded
in large amounts of dust and responsible for the ionizing photons producing 
the emission lines) much more than older, less extincted stellar populations
producing the continuum underlying the lines: the net result 
is a low EW ratio. Additional evidence for a selective
extinction will be discussed in Sec.~3.

\begin{figure}[b]
\vspace{-1.6in}
\begin{center}
\includegraphics[width=.7\textwidth]{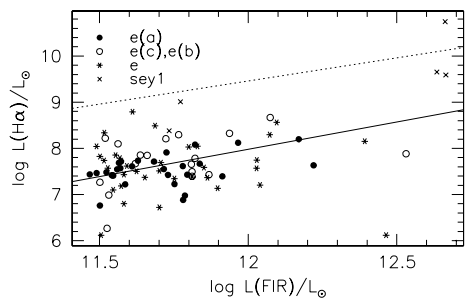}
\end{center}
\vspace{-0.4in}
\caption[]{FIR and $\rm H\alpha$ luminosities in solar units.
No dust correction has been applied.
The best fit to the datapoints is shown as a solid line.
The relation found by Devereux \& Young (1990) for a sample of 
field spiral galaxies in the local Universe 
is extrapolated to the FIR luminosities of the present dataset 
and is shown as a dotted line.
"e" spectra are those with at least one detected emission line,
but $\rm H\delta$ unmeasurable.}
\label{eps1}
\end{figure}
As in all FIR luminous samples, there is a deficiency of $\rm H\alpha$
luminosity at a given FIR luminosity as compared to optically selected
spirals (Fig.~2) and this scarcity of $\rm H\alpha$ flux translates
directly into a difference in the SFR estimate: even applying 
to the $\rm H\alpha$ flux the dust 
correction derived from the Balmer decrement is not sufficient
to reconcile the two estimates of current star formation
obtained from the FIR and $\rm H\alpha$ luminosities (Fig.~3).
After dust correction, the $\rm SFR_{H\alpha}$ is still a factor 
of $\sim 3$ lower than the $\rm SFR_{FIR}$; it is unlikely this discrepancy 
is entirely due to the limited aperture of the spectroscopic slit 
and I will come back to this point in Sec.~3.

\begin{figure}[b]
\vspace{-1.4in}
\begin{center}
\includegraphics[width=.8\textwidth]{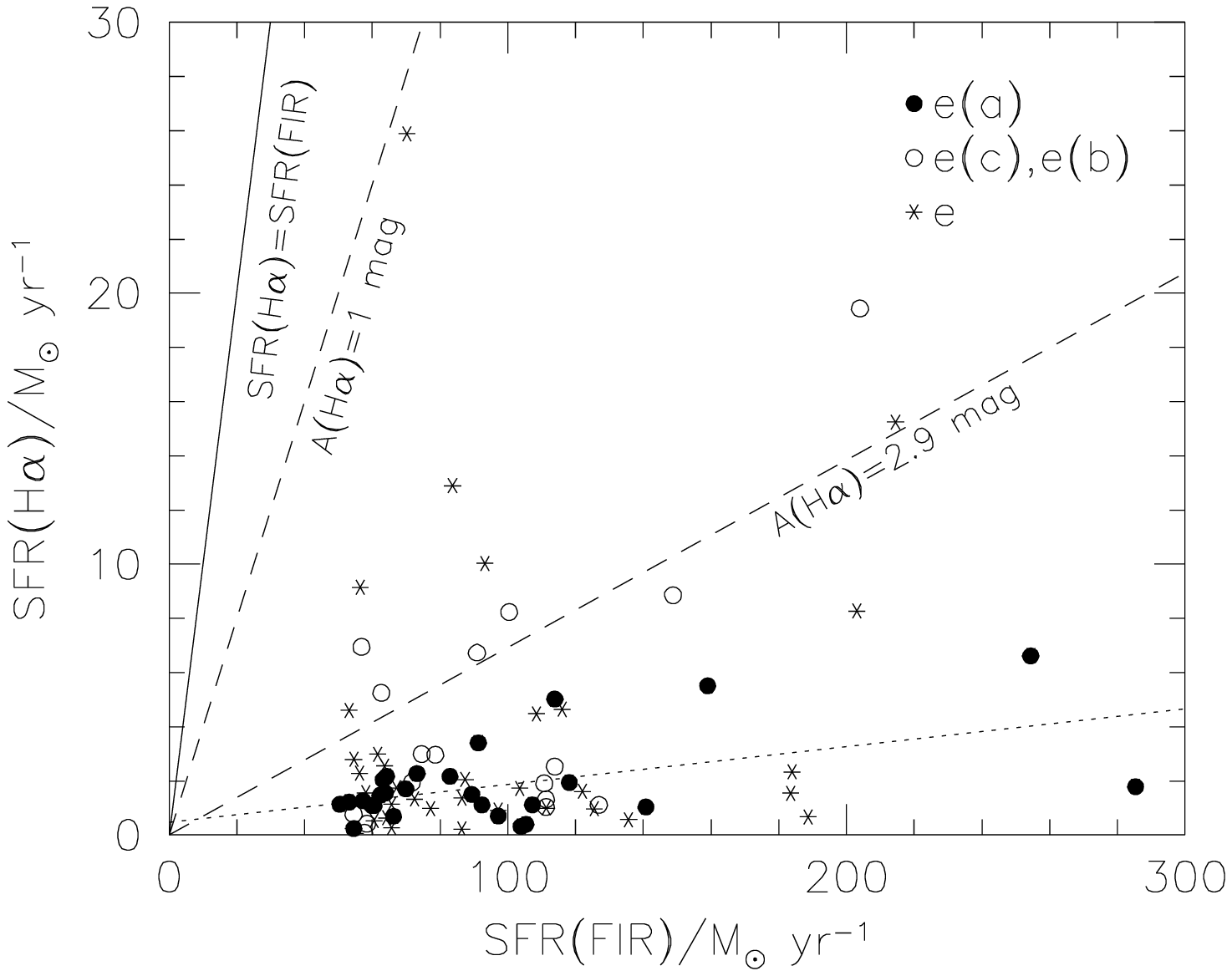}
\end{center}
\vspace{-0.4in}
\caption[]{The SFR derived from the observed $\rm H\alpha$ flux versus the 
FIR--based estimate.
The dotted line is the fit to the e(a) population.
The solid line shows the relation $SFR_{\rm H\alpha}=SFR_{FIR}$,
and the dashed lines are found for 1 mag extinction
at $\rm H\alpha$ (average extinction in nearby spirals)
and for the 
average extinction in e(a)'s in W98 sample as determined by the
Balmer decrement (E(B-V)=1.1, A($\rm H\alpha)=2.9$ mag)}
\label{eps1}
\end{figure}

\begin{table*}
{\scriptsize
\begin{center}
\vspace{0.1cm}
\centerline{\scriptsize \sc Fraction of galaxies and E(B-V) 
as a function of the spectral class.}
\vspace{0.3cm}
\begin{tabular}{lccl}
\hline
\noalign{\smallskip}
Class & \% & Median E(B-V) & Comments (see PW00 for details)\\
\hline
\noalign{\smallskip}
e(a) & 0.56$\pm$0.10 & 1.11 & Strong Balmer absorption plus emission \\
e(c) & 0.25$\pm$0.07 & 0.68 & Weak/moderate Balmer absorption plus emission \\
e(b) & 0.10$\pm$0.04 & 0.62 & Strong emission (EW([O{\sc ii}])$>40$ \AA)\\
sey1 & 0.10$\pm$0.04 & -- & Seyfert1 from broad hydrogen lines in emission \\
\noalign{\smallskip}
\noalign{\hrule}
\noalign{\smallskip}
\end{tabular}
\end{center}
}
\vspace*{-0.8cm}
\end{table*}

\begin{figure}[b]
\vspace{-1.6in}
\begin{center}
\includegraphics[width=.7\textwidth]{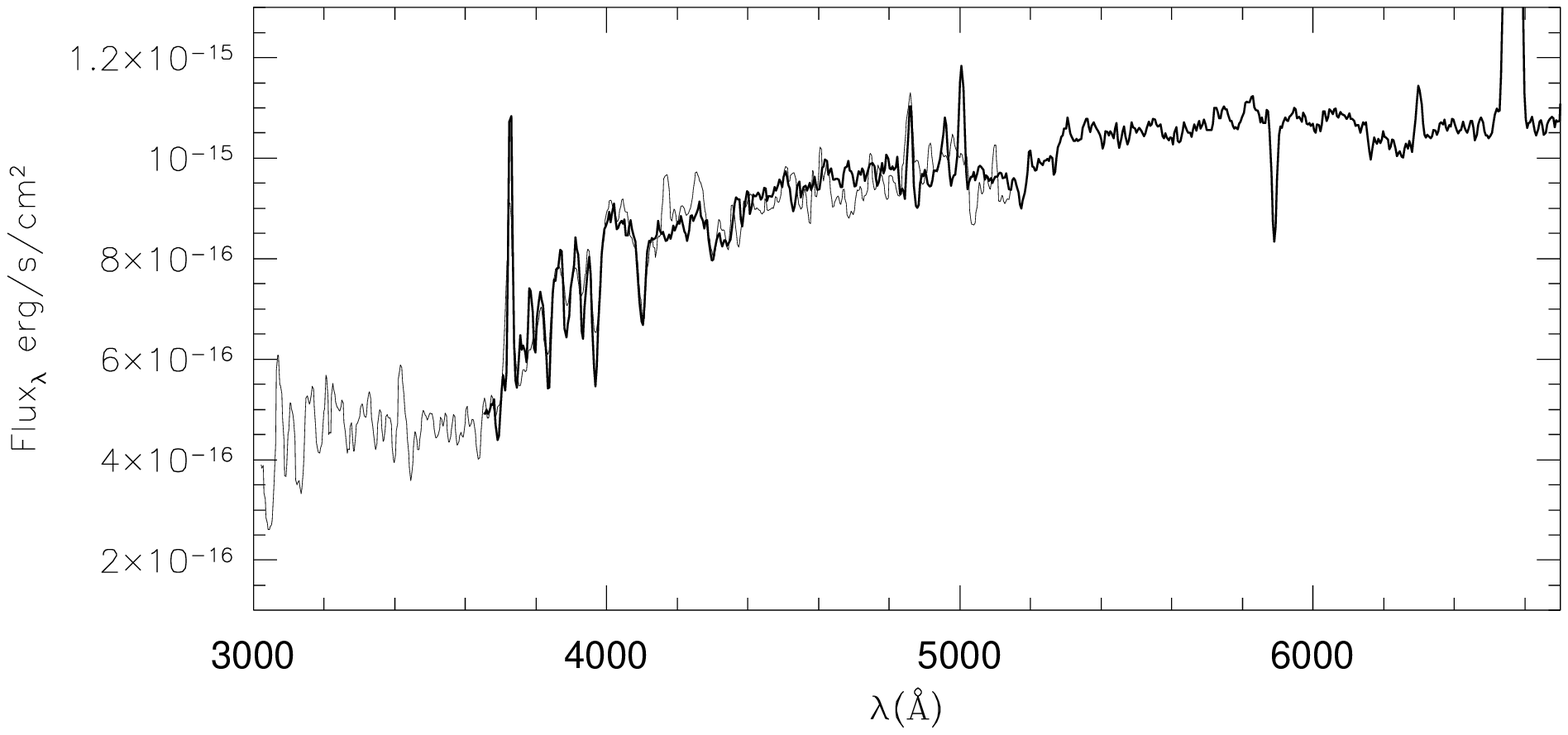}
\end{center}
\vspace{-0.4in}
\caption[]{A rest-frame comparison of the average 
e(a) spectrum of Very Luminous Infrared galaxies  (thick line) and 
the spectrum of the ISO starburst 
galaxies at $<z>\sim 0.5$ (thin line). 
The latter is taken from Fig.~12 in Flores et al. (1999) and has been 
normalized to the VLIRGs spectrum over the wavelength range in common.
This is the average spectrum of 5 ISO-detected galaxies at $z<0.7$ whose 
spectral energy distributions at visible, near-IR, MIR and radio wavelengths
resemble those 
of highly reddened starbursts in the local universe.
The spectral resolution is 10 \AA $\,$
(Wu) and 40 \AA $\,$ (Flores).}
\label{eps1}
\end{figure}
The analysis presented so far refers to a well defined sample of IRAS
luminous galaxies, but the e(a) phenomenon appears to be 
widespread among IR luminous
galaxies at any redshift (see PW00 for a census of the e(a) occurrence).
As an example, Fig.~4 shows that the average e(a) spectrum of Wu's sample
is very similar to the average spectrum of distant starburst galaxies
detected at 15 micron by ISO, both in the continuum shape
and in the strength of the [O{\sc ii}], $\rm H\delta$ and $\rm H\beta$
lines. 

\section{Physical origin: modelling e(a) spectra}

A selective dust extinction has been proposed as the physical origin
of the e(a) spectra (P99, PW00): 
dust obscuration affects the youngest stellar generations more than the older
stars. This is expected to explain both the observed [O{\sc ii}]/$\rm H\alpha$
ratio, as discussed in the previous section, and the $\rm H\delta$ strength
because the stellar populations responsible for this line (with ages a few 
times $10^7$ -- 1.5 $10^9$ yr) have had time to drift away from or disperse
the dusty molecular clouds where they were born and their emission can dominate
the integrated spectrum at 4000 \AA $\,$ if younger stars are more heavily 
obscured. Furthermore, a 
selective extinction appears to be the most plausible explanation
for the fact that, even within the same spectrum, different values of 
extinction are measured depending on the spectral region/feature used 
to estimate it: for example, 
the dust attenuation of the UV/optical stellar continuum
is often measured to be lower than the obscuration of the emission lines
(see PW00 for a reference list).

In order to verify whether the hypothesis of selective extinction
in a starburst galaxy can indeed account for the e(a) spectrum, 
a simplified spectrophotometric
model, including only 10 stellar populations, has been
developed (Poggianti, Bressan \& Franceschini 2001). 
This represents the minimum set of stellar populations
that are known to be essential because they affect the spectral features
that we wish to reproduce: four young generations 
($10^6, 3 \cdot 10^6, 8 \cdot 10^6,
10^7$ yr) responsible for the ionizing photons that produce the 
emission lines;
five intermediate populations ($5 \cdot 10^7, 10^8, 3 \cdot10^8, 5 
\cdot 10^8, 10^9$ yr) with the strongest Balmer lines in 
absorption, and older stars modelled as a constant star formation rate 
between 1 and 12 Gyr before the moment of the observation
which can give a significant contribution to the continuum.
The spectrum of each stellar generation is found from a spectrophotometric 
model that includes both the stellar and the nebular contribution
(Barbaro \& Poggianti 1997) and it is extincted with its own extinction 
value -- that is allowed to vary from a stellar population to another --
assuming a dust screen with a standard Galactic extinction law
and a Salpeter IMF between 0.1 and 100 $M_{\odot}$.
The results of this model are compared with the average e(a) spectrum
of the VLIRGs discussed in Sec.~2; the quality of the fit is assessed
considering the differences between the model and the observed spectrum
in the equivalent widths of four lines ([OII]$\lambda$3727,H$\delta$, 
H$\beta$ and H$\alpha$) 
and the continuum flux in eight spectral windows in the range
3770--6460 \AA.
\begin{figure}[b]
\vspace{-1.0in}
\begin{center}
\includegraphics[width=1.1\textwidth]{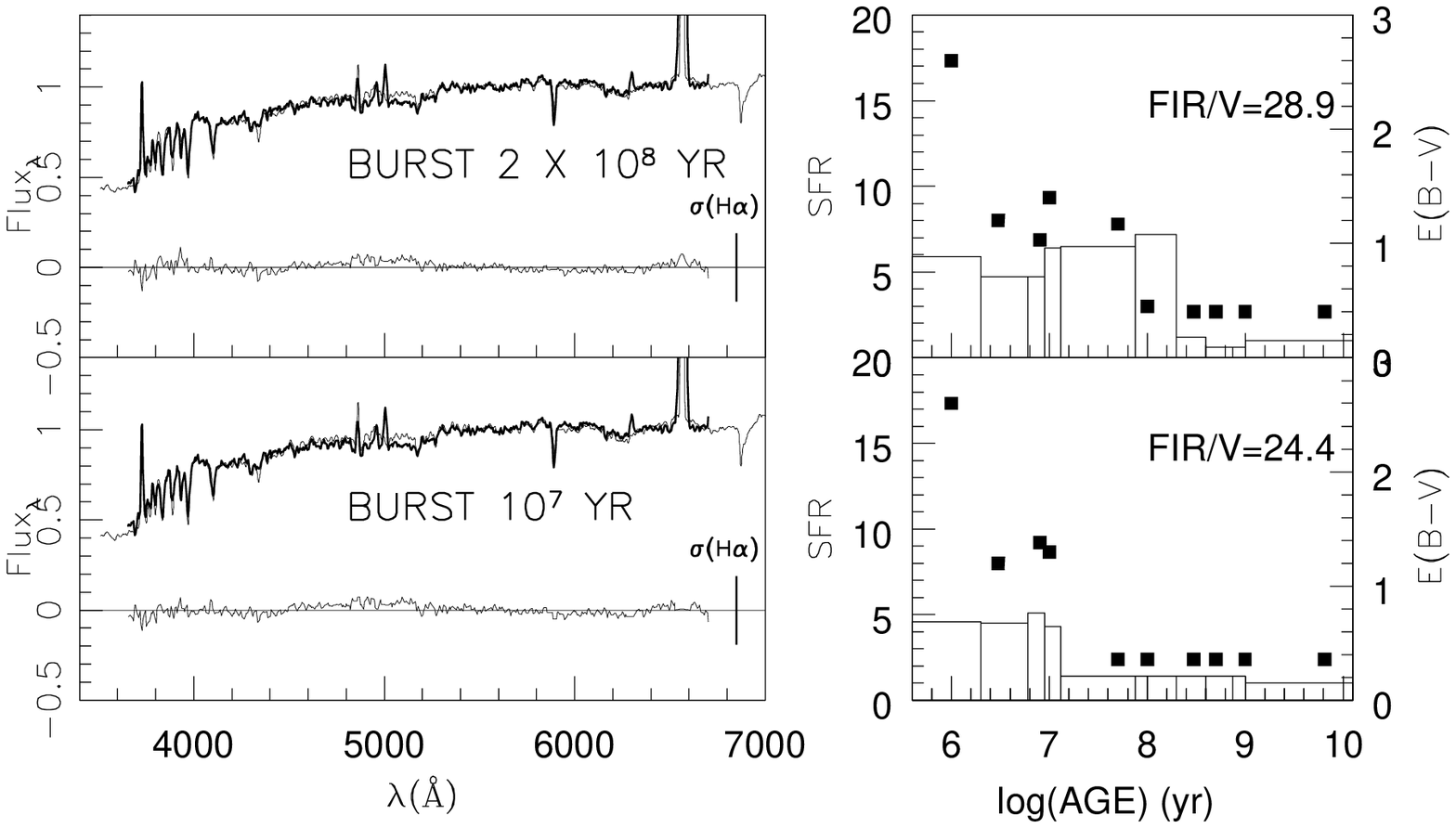}
\end{center}
\vspace{-2.4in}
\caption[]{{\em Left:} Comparison between the average
observed spectrum of the e(a) galaxies (thick line) and the spectrum
of two starburst models (thin lines), normalized at 5500 \AA. 
The emission lines included in the models are those of the Balmer 
series and the [O{\sc ii}] line.
The difference between the model and the observed spectrum is also shown
(at Flux$\sim 0$).
The vertical segment on the right side represents the $1 \sigma$ error in
$\rm H\alpha$. {\em Right:} SFR 
(histogram, normalized to =1 in the old population) and
E(B-V) (dots) of the models whose spectrum is shown in the left panel.
The observed FIR/V ratio is =88.0.}
\label{eps1}
\end{figure}
On the basis of our models we find the following results:

1) the e(a) spectrum is consistent with the 
starburst/selective extinction scenario.
The upper left panel of Fig.~5 presents the comparison between the average
e(a) spectrum and a model of a burst that began $2 \times 10^8$ yr ago;
the extinction of the starburst populations is significantly higher
than that of the older generations (see the right panel in Fig.~5). 
The fit is remarkable, both
as far as the line equivalent widths and the continuum are concerned.
\footnote{There is a small discrepancy around 5000 \AA $\,$
but this is only at the 4\% level.}

2) The model described above 
can only account for about 1/3 of the observed FIR luminosity.
Different effects can contribute to this discrepancy:

a) Slit aperture effects i.e. a mismatch in the galactic area sampled
by the optical spectrum (central 2 kpc) and by IRAS (integrated).
Given that the IR emissivity is usually concentrated in the central
regions of luminous infrared galaxies, it is hard to envisage how
slit effects can account for the discrepancy between the modelled and
the observed IR flux.

b) Dusty starburst galaxies often have 
star forming regions which are completely obscured
at optical wavelengths, hence give no contribution to the spectrum but
produce a significant fraction of the FIR luminosity (e.g. Mirabel 
et al. 1998). The observed FIR/V ratio can be reproduced by starburst models
with regions that are highly obscured by dust with an E(B-V) even greater
than in the models of Fig.~5.

3) No strong constraint can be placed on the burst duration. Models
with a starburst
as short as a few times $10^6$ yr and as long as $10^9$ yr are able to fit
the observed e(a) spectrum as long as:
a) the youngest populations are highly extincted and
b) the old population provides a contribution but does not overwhelm
the intermediate age contribution at 4000 \AA.
As an example, the lower left panel in Fig.~5 shows the excellent fit obtained
with a current burst that began $\sim 10^7$ yr ago. 
In the case of short bursts ($< 5 \times 10^7$ yr), 
the strong $\rm H\delta$ line is not
produced by the stars born during the starburst event, but by the previous 
stellar populations.

4) We also considered a family of "post-starburst models" with extinction, 
assuming 
a galaxy is seen {\em after} a strong starburst phase when a small amount of
residual star formation activity is still ongoing. This type of models --
besides accounting for not more than $\sim 1/10$ of the observed FIR 
luminosity -- fail to reproduce simultaneously the $\rm H\alpha$
and the $\rm H\beta$ lines, either underestimating the former or
overestimating the latter.

\end{document}